\documentstyle[aps,prl,psfig,epsfig,floats]{revtex}
\begin{document}
\draft
\twocolumn[\hsize\textwidth\columnwidth\hsize
\csname @twocolumnfalse\endcsname
\title{Electronic theory for superconductivity in Sr$_2$RuO$_4$:\\
triplet pairing due to spin-fluctuation exchange}
\author{I. Eremin$^{1,2}$, D. Manske$^1$, C. Joas$^1$, and K.H. Bennemann$^1$}
\address{$^1$Institut f\"ur Theoretische Physik, Freie Universit\"at 
Berlin, D-14195 Berlin, Germany}
\address{$^2$Physics Department, Kazan State University, 420008 Kazan, Russia}
\date{February 6, 2001}
\maketitle
\begin{abstract}
  Using a two-dimensional Hubbard Hamiltonian for the three electronic 
  bands crossing the Fermi
  level in Sr$_2$RuO$_4$ we calculate the band structure and spin
  susceptibility $\chi({\bf q}, \omega)$ in quantitative agreement
  with nuclear magnetic resonance (NMR) and inelastic neutron
  scattering (INS) experiments. The susceptibility has two peaks at
  {\bf Q}$_i = (2\pi/3, 2\pi/3)$ due to the nesting Fermi surface 
  properties and at
  {\bf q}$_i = (0.6\pi, 0)$ due to the tendency towards
  ferromagnetism. 
  Applying spin-fluctuation exchange theory as in 
  layered cuprates we determine from $\chi({\bf q}, \omega)$, 
  electronic dispersions, and Fermi
  surface topology that superconductivity in Sr$_2$RuO$_4$ consists of
  triplet pairing. Combining the Fermi
  surface topology and the results for $\chi({\bf q}, \omega)$ we can
  exclude $s-$ and $d-$wave symmetry for the superconducting order parameter.
  Furthermore, within our analysis and approximations we find 
  that $f$-wave symmetry is slightly favored over $p$-wave symmetry due to 
  the nesting properties of the Fermi surface.
\end{abstract}
]
\narrowtext 
The novel spin-triplet superconductivity with $T_c$=1.5K
observed recently in layered Sr$_2$RuO$_4$ seems to be a new example
of unconventional superconductivity\cite{maeno}. The presence of
incommensurate antiferromagnetic and ferromagnetic spin fluctuations
confirmed recently by inelastic neutron scattering (INS)\cite{sidis}
and NMR $^{17}$O Knight shift\cite{imai}, respectively, suggests a
pairing mechanism for Cooper-pairs due to spin fluctuations.  This is
further supported by the observed non $s$-wave symmetry of the order parameter.
Likely Sr$_2$RuO$_4$ is another example of spin
fluctuations induced superconductivity.  This makes the theoretical
investigation of ruthenates very interesting. NMR\cite{kitaoka} and
polarized neutron scattering \cite{duffy} measurements indicate
spin-triplet state Cooper-pairing.  In analogy to $^3$He this led
theorists to conclude that $p$-wave superconductivity is present
\cite{sigrist}.  However, by fitting the specific heat and the
ultrasound attenuation Dahm {\it et al.} doubted $p$-wave
superconductivity\cite{maki} and propose an $f$-wave symmetry of the
superconducting order parameter. A similar conclusion was drawn in
Ref. \onlinecite{hasegawa}.  Recently it has been reported that also
thermal conductivity measurements are most consistent with $f$-wave
symmetry\cite{izawa}.

Clearly, it is important to analyze more definitely the origin of 
superconductivity, triplet pairing and also the symmetry of
the order parameter on a basis of an electronic calculation. This is
difficult, since there are three Ru$^{4+}$ t$_{2g}$ bands that cross
the Fermi level with $\approx$2/3-filling of every band in
Sr$_2$RuO$_4$. The hybridization
between all three bands seems to cause a single $T_c$. All bands cross
the Fermi level and hence, the cross-susceptibilities are not small and play an
important role. In view of these facts the
previous theoretical analysis of the gap symmetries and competition
between $p$ and $d$-wave superconductivity\cite{mazin,kuroki,ogata}
must be re-examined and it is necessary to determine superconductivity
within an electronic theory and to derive the symmetry of the order
parameter from general arguments.

In this letter we present an electronic theory which takes into
account the hybridization between all three bands. We calculate the
Fermi surface (FS), energy dispersion and the spin susceptibility
$\chi$ including {\it all} cross-susceptibilities. Then, we analyze the
pairing interaction mediated by the spin fluctuations exchange in
Sr$_2$RuO$_4$. Analyzing experimental results for the $^{17}$O Knight
shift and INS data as well as the FS observed by
Angle-Resolved-Photoemission-Spectroscopy (ARPES)\cite{shen} we obtain
values for the hopping integrals and effective Coulomb repulsion $U$.
Taking this as an input into the pairing interaction we analyze the
$p$-, $d$- and $f$-wave superconducting gap symmetries. The 
delicate competition between weak ferromagnetic spin fluctuations 
and relatively strong
incommensurate antiferromagnetic spin fluctuations due to nesting of
the FS cause triplet Cooper-pairing. We get 
singulet d$_{x^2-y^2}$-wave symmetry is energetically less favorable.

We start from the two-dimensional three-band Hubbard Hamiltonian
\begin{equation}
H=\sum_{{\bf k}, \sigma} \sum_{\alpha} t_{{\bf k} \alpha} 
a_{{\bf k}, \alpha \sigma}^{+} a_{{\bf k}, \alpha \sigma} +
 \sum_{i,\alpha} U_{\alpha} \, n_{i \alpha \uparrow} n_{i \alpha \downarrow},
\label{hamilt}
\end{equation}
where $a_{{\bf k}, \alpha \sigma}$ is the Fourier transform of the
annihilation operator for the $d_{\alpha}$ orbital electrons ($\alpha
= xy, yz, zx$) and $U_{\alpha}$ is an effective on-site Coulomb
repulsion. $t_{{\bf k}\alpha}$ denotes the energy dispersions of the
tight-bindings bands calculated as follows: $t_{{\bf k}\alpha} = -
\epsilon_0 -2t_x \cos k_x - 2t_y \cos k_y +4t' \cos k_x \cos k_y$. In
accordance with experimental measurements of the Fermi surface and
energy dispersions we choose the values for the parameter set
($\epsilon_0, t_x, t_y, t'$) as (0.5, 0.42, 0.44, 0.14), (0.23, 0.31,
0.055, 0.01), and (0.24, 0.045, 0.31, 0.01)eV for $d_{xy}$-,
$d_{zx}$-, and $d_{yz}$-orbitals\cite{shen}.  The analysis of de
Haas-van Alphen experiments\cite{dvha} shows a substantial
hybridization between $xz$- and $yz$- orbitals about $t_{\perp}$ = 0.1
eV, but not with the $xy$-orbital\cite{morr}. However, the observation
of a single $T_c$ implies the coupling between all three bands.
Therefore, we choose a weak hybridization $t_{hyb}=0.01\mbox{eV}$
(hybridization between $xy$- and $xz$-,$yz$-orbitals) $\ll t_{\perp}$.
Note, even such a weak hybridization transfers the nesting
properties to the $xy$-orbital.
In the inset of Fig. 1 we show the resultant energy dispersions of the
obtained hole-like $\alpha$-band and electron-like $\beta$- and
$\gamma$-bands after hybridization. Due to the small value of
hybridization between $xy$ and $yz$, $xz$ orbitals the dispersion
curves and resulting FS (see also Fig. 4) look quite similar to the
non-hybridized ones\cite{takimoto}.  However, the importance of
hybridization between these orbitals for the spin susceptibility,
$\chi ({\bf q}, \omega)$, will be immediately seen from the analysis 
of the latter.
The susceptibility is given by:
\begin{equation}
\chi_{0}^{ij}({\bf q},\omega) = \frac{1}{N}\sum_{\bf k} 
\frac{f(\epsilon^{i}_{{\bf k}, \gamma}) -f(\epsilon^{j}_{{\bf k+q},\gamma})}
{\epsilon^{i}_{{\bf k+q},\gamma}-\epsilon^{j}_{{\bf k},\gamma} + 
\omega + i0^+}
\quad,
\label{lindhard}
\end{equation}  
where $f(\epsilon)$ is the Fermi function and $\epsilon^{i}_k$ is the
energy dispersion of the $\alpha$, $\beta$, and $\gamma$ band\cite{matrix}.
\begin{figure}[t]
%
\vspace{-1.5cm}
\centerline{\epsfig{clip=,file=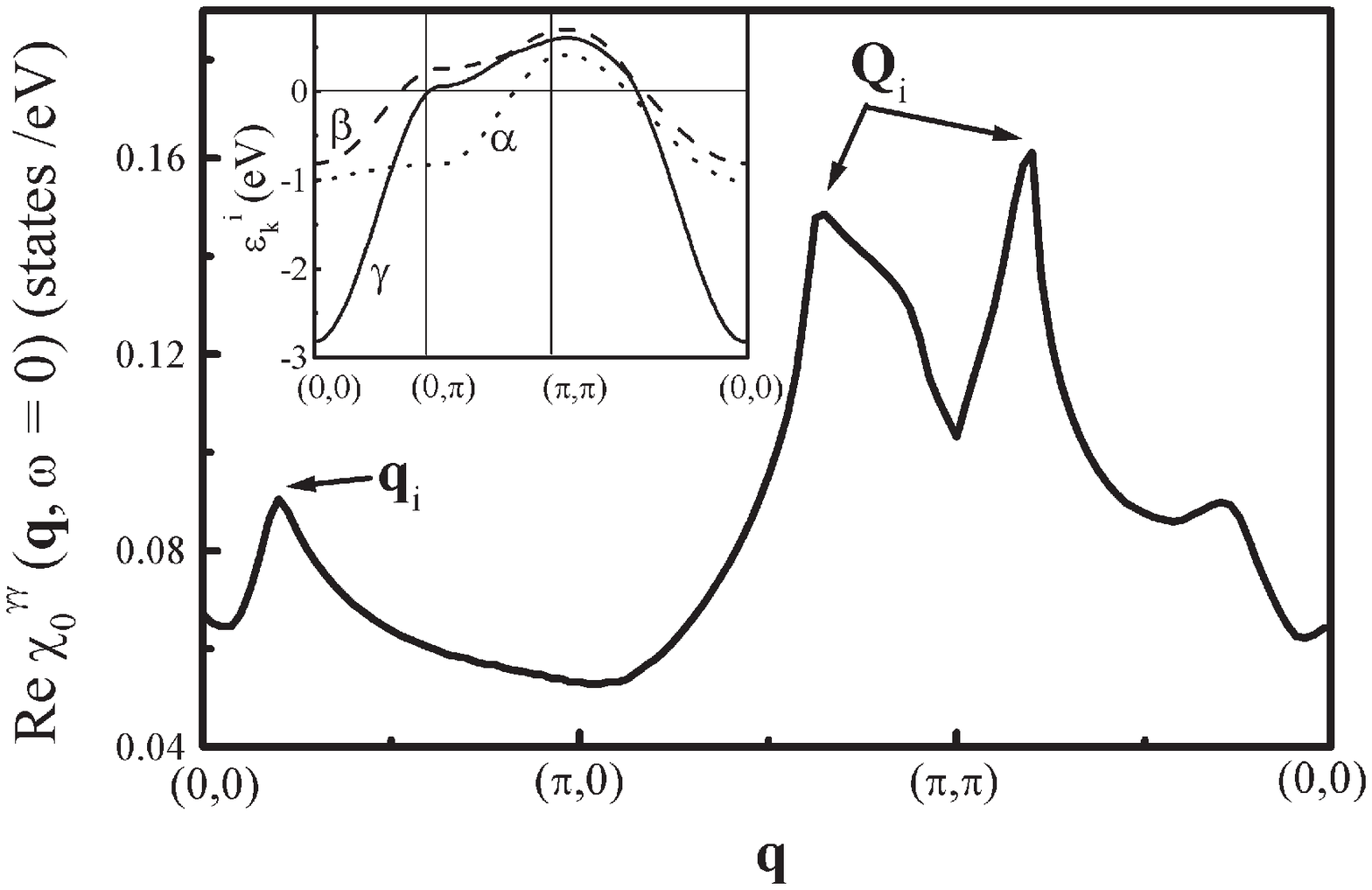,width=7.7cm,angle=0}}
\vspace{0.5ex}
\caption{Calculated susceptibility 
  Re $\chi_{0}^{\gamma\gamma} ({\bf q}, \omega = 0)$ obtained from
  electronic calculations using the hybridized bands.  The wave
  vectors {\bf Q}$_i = (\frac{2\pi}{3}, \frac{2\pi}{3})$ and {\bf
    q}$_i$ $\approx$ $(0.2\pi, 0)$ reflect nesting transferred from
  $\alpha$ and $\beta$ bands and the original tendency of the $\gamma$
  band towards ferromagnetism, respectively. These wave vectors play
  the most important role for Cooper-pairing. The inset shows results
  for the energies $\epsilon_{k}^{i}$ of the hybridized bands ($i=\alpha$,
  $\beta$, $\gamma$).  }
\label{fig1}
\end{figure}

In Fig. 1 we show the momentum dependence of the real part of
$\chi_0^{\gamma\gamma}$. While hybridization between bands does not
affect much the energy dispersion, it changes significantly the
susceptibility of the $\gamma$-band. In particular, the nesting
properties of $xz$- and $yz$-orbitals reflected by the peak at ${\bf
  Q}_i=(2\pi/3, 2\pi/3)$ in $\chi_0^{\gamma\gamma}$ are caused by the
hybridization between $xz$, $yz$ and $xy$ bands. Note, without taking
into account the hybridization one would not get the peak at ${\bf
  Q}_i$ in the $\gamma$-band, but only a broad hump as discussed
earlier\cite{takimoto}. The small peak at ${\bf q}_i\approx(0.2\pi,0)$
is due to the original tendency towards ferromagnetism of the
$xy$-band and is not affected by the hybridization.
 
Our results for $\chi({\bf q}, \omega)$ will have important consequences 
and were obtained in contrast to previous studies from observed band 
dispersions. The susceptibility matrix [$\chi^{ij}$] is calculated where 
$i$, $j$ refers to the hybridized bands. As already 
mentioned the non-diagonalized 
susceptibilities are not small and thus cannot be neglected.
After diagonalizing the matrix $[\chi_0^{ij}]$ we get within RPA 
with an effective $U({\bf q})$ for $\chi$:
\begin{equation}
\chi({\bf q}, \omega) = \frac{\chi_{0}({\bf q}, \omega)}
{1-U({\bf q})\chi_{0}({\bf q}, \omega)}
\quad,
\label{RPA}
\end{equation} 
where now $\chi_0({\bf q}, \omega) = \sum_{i'}\chi^{i'}_0({\bf q},
\omega)$.  Here, $\chi^{i'}_0({\bf q}, \omega)$ ($i'=\alpha'$,
$\beta'$, and $\gamma'$) are the diagonal elements of the diagonalized
matrix $[\chi_0^{ij}]$. This obtained susceptibility characterizes the normal
state magnetic properties of Sr$_2$RuO$_4$. Its spin fluctuations are
given by $\chi({\bf q}, \omega)$ with peaks at {\bf Q}$_i$ and
{\bf q}$_i$.  These wave vectors are important for determining the
symmetry of the superconducting order parameter.
\begin{figure}[t]
\vspace{-1.5cm}
\centerline{\epsfig{clip=,file=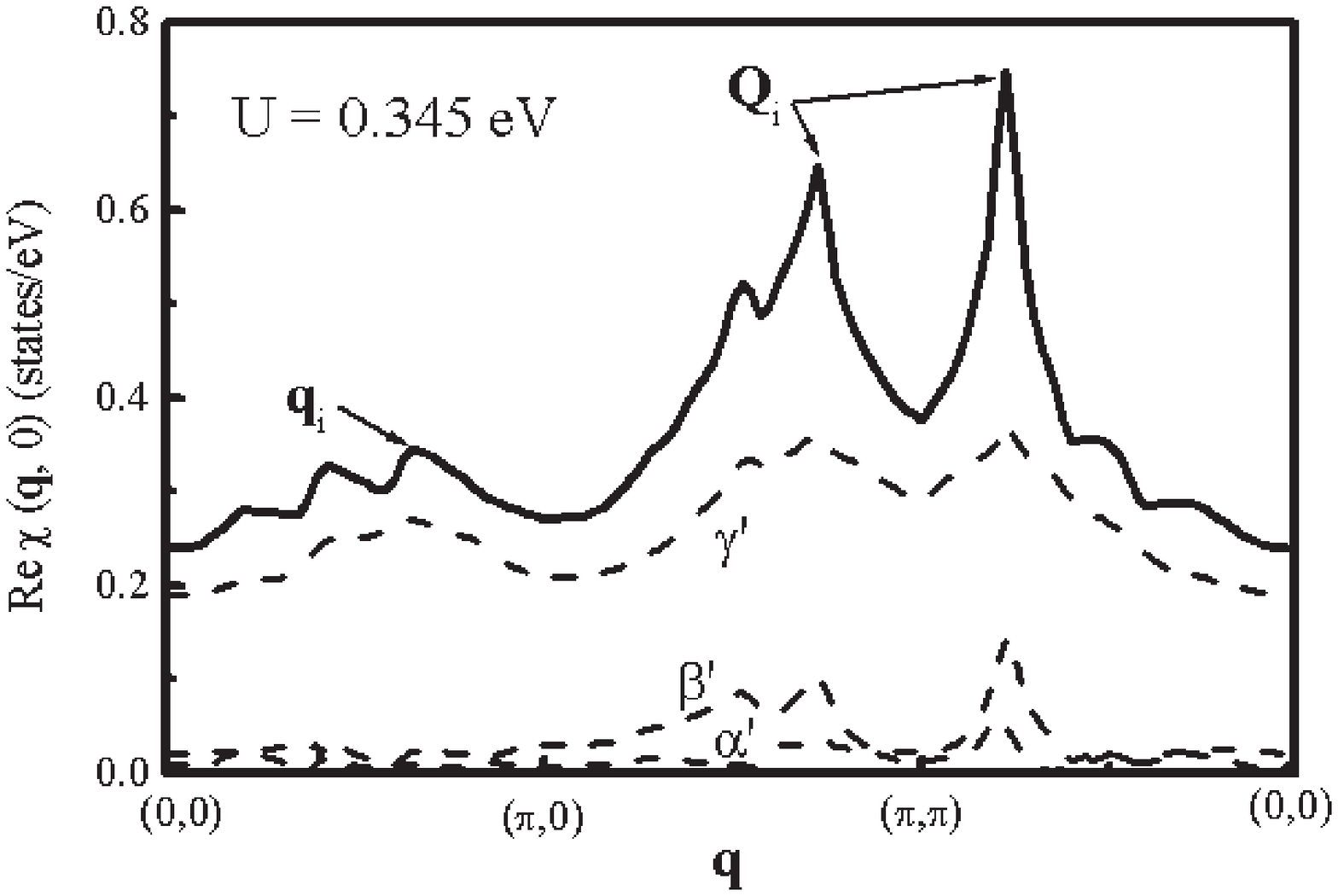,width=7.7cm,angle=0}}
\vspace{0.5ex}
\caption{Results are shown for the static susceptibility at T = 50K.
  The solid curve refers to the total susceptibility within RPA
  resulting from the partial susceptibilities Re $\chi_{0}^{i'}$
  ($i=\alpha'$, $\beta'$, $\gamma'$) shown by the dashed curves. The
  $\chi_{0}^{i'}$ refer to the diagonal elements of the diagonalized
  matrix $[\chi_0^{ij}]$.  Note, the pairing wave vectors ({\bf
    Q}$_i$, {\bf q}$_i$) are nearly the same as in Fig. 1. Note, the
  smallness of $\chi_{0}^{\alpha'}$ and $\chi_0^{\beta'}$ as compared
  to $\chi_{0}^{\gamma'}$.
  }
\label{fig2}
\end{figure}

In Fig. 2 we present the results for $\chi({\bf q}, \omega)$
obtained from Eq. (\ref{RPA}) and the susceptibilities
$\chi_{0}^{i'}({\bf q}, \omega)$ obtained after diagonalization of
$[\chi_{0}^{ij}({\bf q}, \omega)]$.  Here, we approximate $U({\bf q})$
by $U = 0.345$eV which gives agreement with INS\cite{rem}.
Remarkably, the peak in Re $\chi$ at ${\bf Q}_i$ remains nearly the
same as for $\chi_0^{\gamma\gamma}$.  The peak at ${\bf q}_i$ is also
present, but slightly shifted to a larger value. Clearly,
$\chi_0^{\gamma'}$ is much larger than $\chi_0^{\alpha'}$ and
$\chi_0^{\beta'}$. Our calculations also have shown that
cross-susceptibilities $\chi_{0}^{ij}$ ($i\neq j$) cannot be
neglected.
\begin{figure}[t]
\vspace{-1.1cm}
\centerline{\epsfig{clip=,file=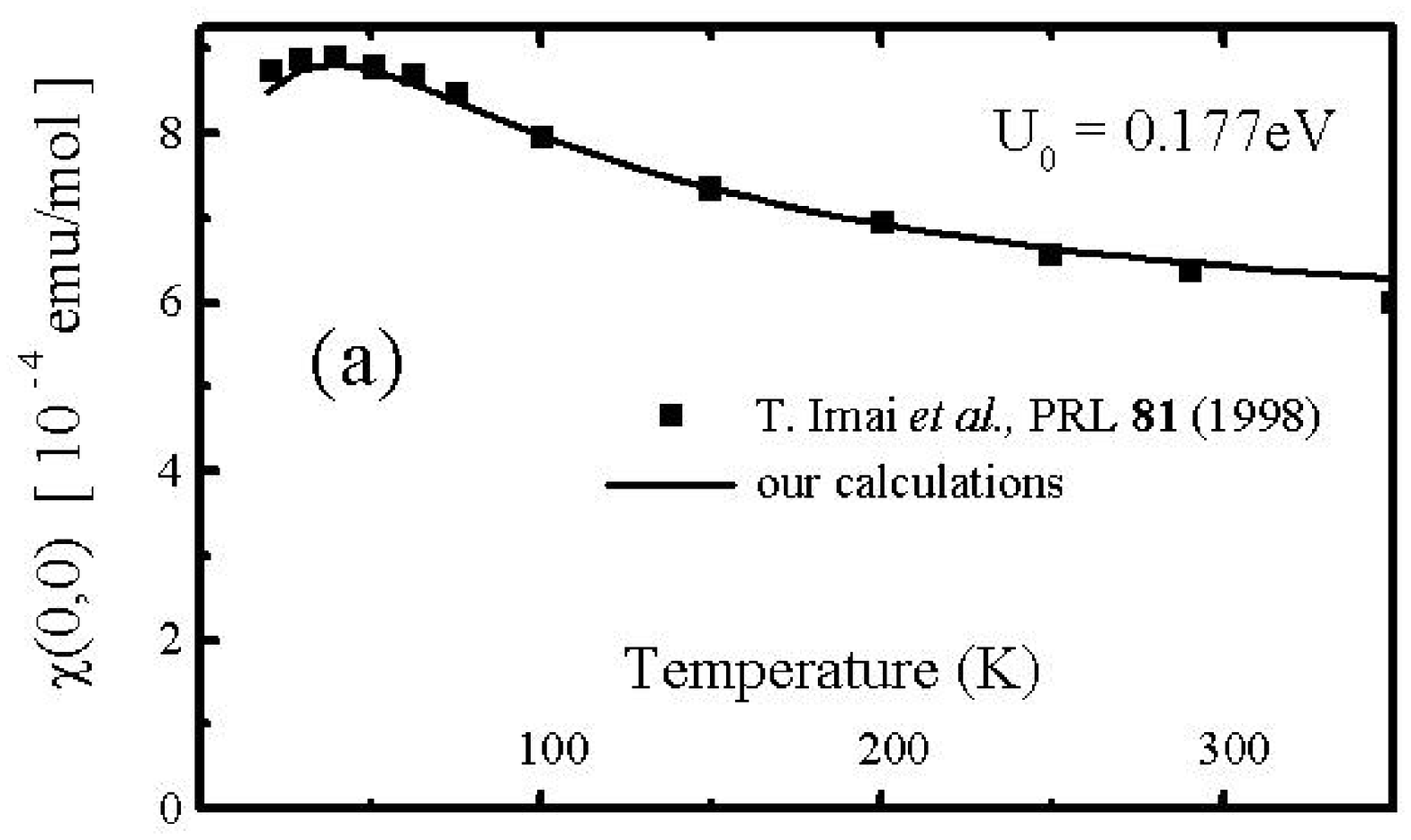,width=7.5cm,angle=0}}
\vspace{-0.75cm}
\centerline{\epsfig{clip=,file=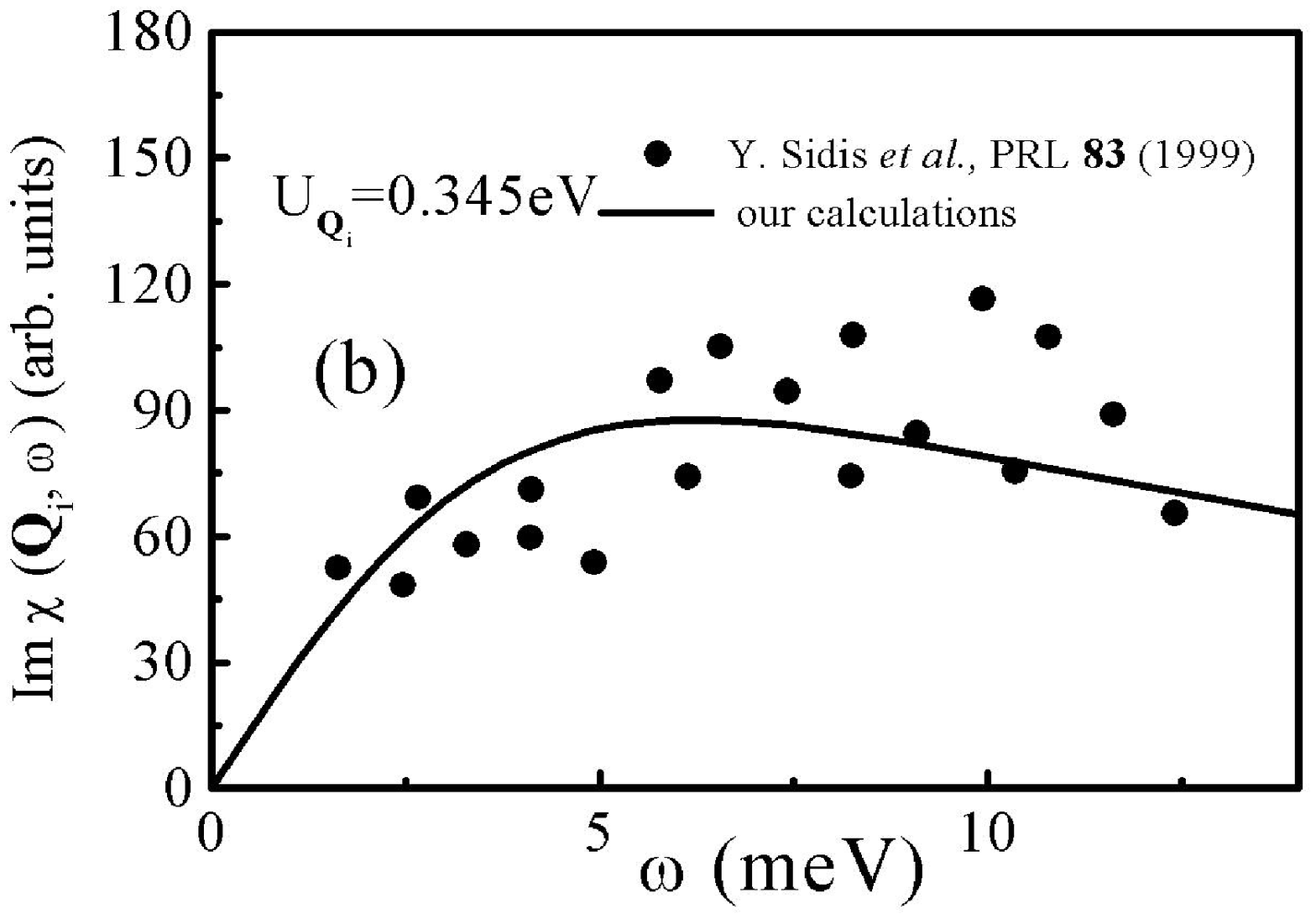,width=7.7cm,angle=0}}
\vspace{0.5ex}
\caption{ (a) Calculated temperature dependence of the uniform spin
  susceptibility with U$_0$=0.177eV is compared with the $^{17}$O
  Knight shift measurements. The peak is due to thermal activation
  involving $\gamma$ and $\alpha$, $\beta$ bands. (b) Calculated
  frequency dependence of Im $\chi({\bf Q}_i, \omega)$ compared
  to INS data using U$_{{\bf Q}_i}$ = 0.345eV.  }
\label{fig3}
\end{figure}

In Fig. \ref{fig3}(a) we compare our calculation of the temperature
dependence of the uniform spin susceptibility $\chi (0,0)$ which
is measured by the $^{17}$O Knight shift\cite{imai}, and in Fig.
\ref{fig3}(b) we compare Im $\chi ({\bf Q}_i, \omega)$ with INS
data\cite{sidis}.  For the calculation of $\chi (0,0)$ we
approximate $U({\bf q})$ by $U(0)=0.177$eV\cite{liebsh} which gives
agreement with Knight shift measurements and is also taken in 
previous calculations.  These comparisons shed light on the validity of 
our results for $\chi({\bf q}, \omega)$. Note, we also take into account
that there are four electrons per three $t_{2g}$-bands that would give
every $\chi_0^{i'}$ an additional weight $4/3$.  Our results are in
fair agreement with experiment that shows a tendency towards  
ferromagnetism\cite{remark}.  The maximum in
$\chi_{RPA} (0,0)$ at about 25K results from thermally activated
changes in the populations of the bands near $E_F$.  In Fig. 3(b) we
compare our results with INS data. In this case we must take 
$U_{{\bf Q}_i}$ = 0.345eV in order to fit $\chi({\bf q}, \omega)$ to 
the peak position and height at $\omega$= 6meV as observed in INS.

While  an uncertainty in the INS data (shown in Fig. \ref{fig3}(b)) 
is present, our results for $\chi({\bf q}, \omega)$ should 
be a useful basis for further calculations. The antiferromagnetic 
spin excitations result in incommensurate 
antiferromagnetic Ru-spin alignment at distances larger than 
nearest neighbors. Hence, if Cooper-pairing involves 
nearest neighboring Ru spins also incommensurate antiferromagnetic 
fluctuations will cause triplet-pairing because neighboring Ru spins 
see also partly a {\it ferromagnetic} environment. 
Note, $\chi({\bf q}, \omega)$ controls the symmetry of the 
superconducting order parameter.

For the analysis of superconductivity in Sr$_2$RuO$_4$ we take
into account that experiment observes non-$s$-wave symmetry of the
order parameter which strongly suggests spin-fluctuation-mediated
Cooper pairing. Assuming the spin-fluctuation-induced pairing it is
possible to analyze the symmetry of the superconducting state from the
gap equation and our calculated results for $\chi({\bf q}, \omega)$
with the pronounced wave vectors at {\bf Q}$_i$ and {\bf q}$_i$.
We get for the gap equation:
\begin{equation}
\Delta^{i}_{\bf k} = 
- \sum_{{\bf k}', j} \lbrack V^{eff}_{\sigma} ({\bf k, k}')\rbrack^{ij} 
\frac{\Delta_{{\bf k}'}^{j}}
{2 E_{{\bf k}'}^{j}}\tanh \Big(\frac 
{E_{{\bf k}'}^{j}}{2k_B T}\Big) 
\quad,
\label{bcs}
\end{equation}
where 
$E_{\bf k}^{i} = \sqrt{\epsilon^{i}_{\bf k} \,^2 +\Delta_{\bf k}^2}$ are 
the energy dispersions of the bands\cite{remeuro} 
and the pairing potential $V^{eff}_{\sigma} ({\bf k, k}')$ is
different for singulet ($\sigma=0$) and triplet ($\sigma=1$) Cooper
pairing.  The eigenvalue analysis of Eq. (4) will yield the symmetry with 
lowest energy. The $\gamma$-band plays the most important role.

For the determination of the
pairing symmetry we follow the analysis by Anderson and Brinkmann for
$^3$He \cite{anderson} and by Scalapino for the
cuprates\cite{scalapino} and use the calculated FS and spin
susceptibility for Sr$_2$RuO$_4$.  For triplet pairing the effective
pairing interaction is ($U\equiv U({\bf Q}_i)$)
\begin{eqnarray}
\lefteqn{V^{eff}_{1} ({\bf k, k}')= - \frac{U^2 \chi_0({\bf k-k}', 0)}
{1- U^2 \chi^{2}_{0}({\bf k-k}', 0)} =} \nonumber\\
& & - \frac{U^2}{2}\Bigg(
\frac{\chi_0({\bf k-k}', 0)}{1- U\chi_0({\bf k-k}', 0)} +  
\frac{\chi_0({\bf k-k}', 0)}{1+ U\chi_0({\bf k-k}', 0)}\Bigg) 
\quad,
\label{triplet}
\end{eqnarray}
and for singulet pairing
\begin{equation}
V^{eff}_{0} ({\bf k, k}') = \frac{U^2\chi_0({\bf k-k}', 0)}
{1- U\chi_0({\bf k-k}', 0)} + \frac{U^3 \chi^2_0({\bf k-k}', 0)}{1-
U^2\chi^2_0({\bf k-k}', 0)}
\quad,
\label{singlet}
\end{equation}
respectively\cite{remark2}.  It is important to note that for E$_F \gg
\Delta_l$ the gap function can be expanded into spherical harmonics
corresponding to the angular momentum $l$ = 1, 2, 3, ... and no
mixture of $\Delta_l$ belonging to different symmetry representations 
can be present if a single $T_c$ is observed.
Therefore, we can exclude the ($p$ + $d$)-wave superconducting state.
Using appropriate symmetry representations\cite{hasegawa} we discuss
the solutions of Eq. (\ref{bcs}) for the $p$, $d$, and $f$-wave
symmetries of the order parameter:
\begin{eqnarray}
\Delta_p({\bf k}) & = & \Delta_0 {\bf \hat{z}}
(\sin k_x + i \sin k_y), \\
\Delta_d({\bf k}) & = & \Delta_0 (\cos k_x -\cos k_y), \\
\Delta_f({\bf k}) & = & \Delta_0 {\bf \hat{z}}
(\cos k_x -\cos k_y)(\sin k_x + i \sin k_y).
\label{symmetry}
\end{eqnarray}
Note, the largest eigenvalue in Eq. (\ref{bcs}) will yield the
superconducting symmetry of $\Delta_l$ in Sr$_2$RuO$_4$. 
Solving Eq. (4) in the first BZ down to 5K, we find $f$-wave symmetry 
slightly favored. As expected $p$- and $f$-wave symmetry Cooper-pairing 
are close in energy ($\lambda_f = 0.76$ $>$ $\lambda_p = 0.51$). 
A more complete 
analysis taking into account also the 
coupling between RuO$_2$ planes and interband $U$ 
might yield a definite answer\cite{footbe}.
Note, 
to obtain a combined energy gain from the antiferromagnetism and 
Cooper-pairing one expects an order parameter with nodes in the RuO$_2$ 
plane and possibly also with respect to the $c$-direction. 

The solutions of Eq. (\ref{bcs}) can be characterized by 
Fig. 4 where we present our
results for the Fermi surface, wave vectors {\bf Q}$_i$ and 
{\bf q}$_i$ and symmetry of the order parameter in Sr$_2$RuO$_4$. 
The
areas with $\Delta_f>0$ and $\Delta_f<0$ are denoted by (+) and (-),
respectively. In a good approximation we linearize Eq.
(\ref{bcs}) in $\Delta_l$, i.e. $E_{{\bf k}'}^{\gamma'} \rightarrow
\epsilon_{{\bf k}'}^{\gamma}$, and safely put $\tanh (\epsilon_{{\bf
    k}'}^{\gamma}/2k_B T)=1$.  Therefore, the main contribution to the
pairing comes from the Fermi level. Note, determining $\Delta_l$ 
for the RuO$_2$-planes it is sufficient to take into account only the 
$\gamma$-band, since only this band has a dispersion in the plane. 
The minus sign in Eq. (\ref{bcs}) is cancelled for
triplet pairing (see Eq. (\ref{triplet})).  Furthermore, the summation
over ${\bf k}'$ in the first BZ is dominated by the contributions due
to ${\bf Q}_i$ and the one due to the background and ${\bf q}_i$.
Thus, we obtain approximately for the $\gamma$-band contribution ($l=f$ 
or $p$)
\begin{eqnarray}
\Delta_l({\bf k}) & \approx & \sum_i \frac{V^{eff}_1 ({\bf Q}_i)}{2
\epsilon_{{\bf k+Q}_i}^{\gamma}} \Delta_l({\bf k+Q}_i) + \nonumber\\
& & {}+\sum_i \frac{V^{eff}_1 ({\bf q}_i)}{2
\epsilon_{{\bf k+q}_i}^{\gamma}} \Delta_l({\bf k+q}_i)
\quad,
\label{simtripl}
\end{eqnarray}
where the sum is over all contributions due to {\bf Q}$_i$ and {\bf
  q}$_i$.  As can be seen from Fig. 4 in the case of $f$-wave symmetry
the wave vector ${\bf q}_i$ in Eq. (\ref{simtripl}) bridges the same
number of portions of the FS with opposite and equal sign.  
Therefore, the second
term in Eq. (\ref{simtripl}) is approximately zero for triplet
pairing.  We see from Fig. 4 that
${\bf Q}_i$ bridges portions of the FS with {\it equal} signs of the
superconducting order parameter. Thus, a solution of Eq.
(\ref{simtripl}) for $\Delta_f$ is indeed possible.
\begin{figure}[t]
\vspace{-0.5cm}
\centerline{\epsfig{clip=,file=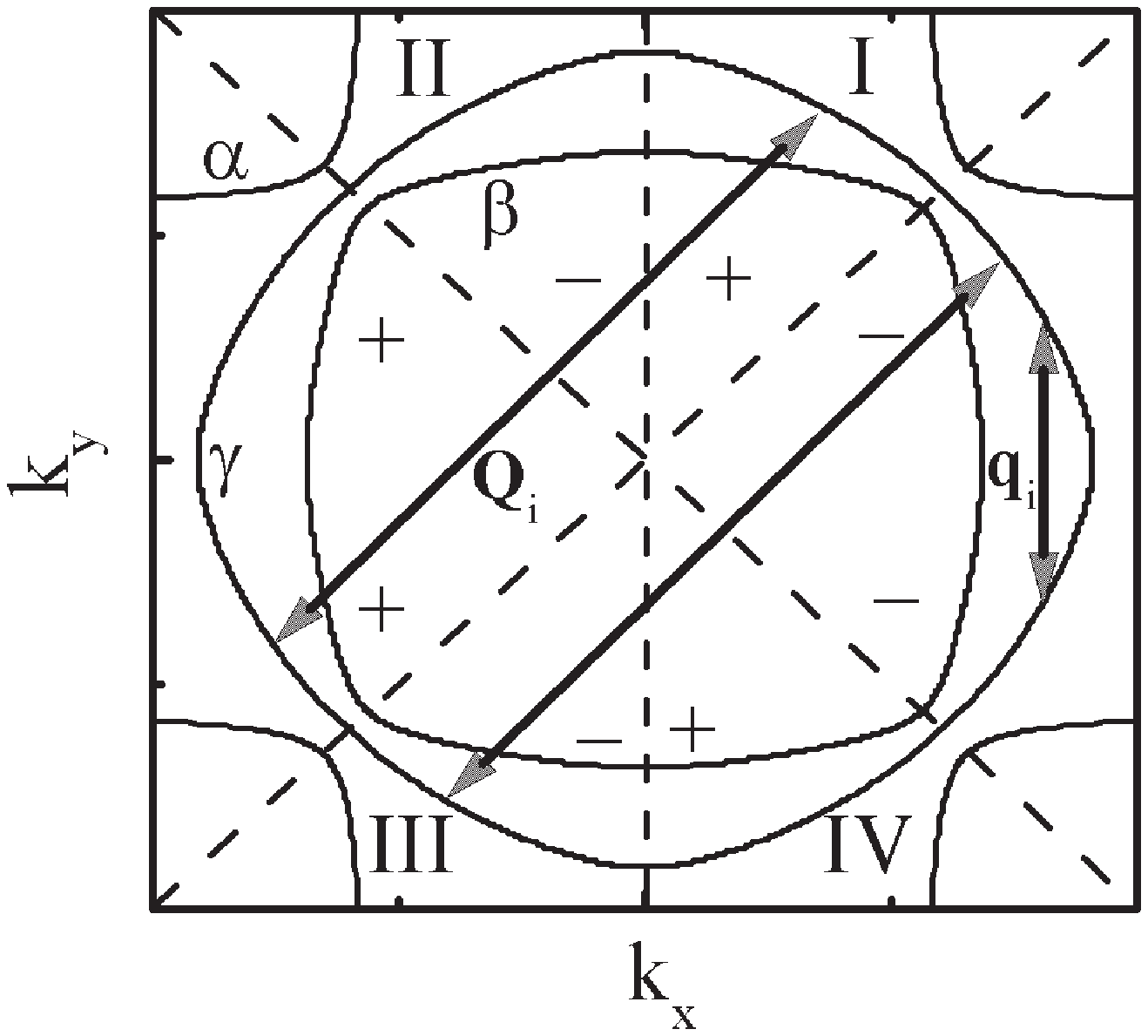,width=7.6cm,angle=0}}
\vspace{0.5ex}
\caption{Symmetry analysis of the order parameter for
  the triplet pairing in the first BZ. $\alpha$, $\beta$, and $\gamma$
  denote the FS of the corresponding hybridized bands. The wave
  vectors {\bf Q}$_i$ and {\bf q}$_i$ are the pronounced wave vectors
  resulting from the susceptibility shown in Fig. 2. These determine
  the symmetry of the order parameter. Also for $f$-wave symmetry 
  the nodes of the real part
  of the order parameter are shown (dashed lines) and the
  regions + ( - ) where the $f$-wave superconducting gap is positive
  (negative). Note, for the real part of the $p$-wave order parameter  
  the node occurs along $k_x $=0.}
\label{fig4}
\end{figure}

In the case of $p$-wave pairing the real part of the order 
parameter has a node only along $k_x =$0 in the $k_x,k_y$-plane. Then, 
using the corresponding Fig.4 
the wave vectors ${\bf Q}_i$ bridge portion of the FS where Re$\Delta_p$ 
has the same or opposite sign. Regarding the ${\bf q}_i$ contributions the 
situation is similar as in the case of the $f$-wave symmetry. Hence, we 
expect for the eigenvalues $\lambda_p \le \lambda_f$ as in the result of 
the algebraic solution of Eq. (4). Note, for increasing nesting Fig. 4 also 
suggests that $f$-wave symmetry is favored more than $p$-wave.
The eigenvalue analysis of the possible solutions $\Delta_f$ and 
$\Delta_p+i\Delta_f$ should increasingly rule out the latter for 
stronger nesting
 
Also using similar arguments we can rule out singulet pairing 
 on the basis of Eq. (\ref {singlet}). 
In particular, assuming $d_{x^2 -y^2}$-symmetry for Sr$_2$RuO$_4$ we
get a change of sign of the order parameter upon crossing the
diagonals of the BZ.  According to Eq. (\ref{bcs}) wave vectors around
{\bf Q}$_i$ connecting areas (+) and (-) contribute constructively to
the pairing. Contributions due to {\bf q}$_i$ and the background
connecting the same sign areas subtract from the pairing 
(see Fig. 4 with nodes at the diagonals for illustration). 
Therefore, we get that the four contributions 
due to {\bf q}$_i$ and the
background cancel the pair-building contribution due to {\bf Q}$_i$.
As a consequence we obtain no $d_{x^2 - y^2}$-wave symmetry.  Note,
this is in contrast to the cuprates where the cancelling contributions
due to {\bf q}$_i$ and the small background are negligible. For the
$d_{xy}$-symmetry where the nodes are along ($\pi$,0) and (0,$\pi$)
directions we can argue similarly and thus exclude this symmetry.

Thus, as a result of the topology of the FS and the spin
susceptibility we get for $p$- and $f$-wave the strongest pairing 
and can definitely exclude $d$-wave pairing. In our approximation 
we find that $f$-wave symmetry pairing is slightly favored
over $p$-wave symmetry in Sr$_2$RuO$_4$.

In view of our Fig. 4 we also remark that while Re $\Delta_f$ exhibits
three line nodes, that can be seen by phase sensitive experiments,
$|\Delta_f|^2$ shows nodes only along the diagonals, as recently found
in measurements of ultrasound attenuation below T$_c$\cite{maeno2}. However, 
note in view of the low eigenvalues for $p$ and $f$-wave symmetries and 
approximations used we cannot definitely conclude that $f$-wave is favored 
over $p$-wave. 

In summary, we show that hybridization between all three bands is
important and transfers the nesting properties of $xz$- and
$yz$-orbitals to the $\gamma$ band in Sr$_2$RuO$_4$.  Taking into
account all cross-susceptibilities we successfully explain the
$^{17}$O Knight shift and INS data. Most importantly, 
we calculate $\chi({\bf q}, \omega)$ and show on the basis of the
Fermi surface topology and the calculated spin susceptibility
$\chi({\bf q}, \omega)$ that triplet pairing is present
in Sr$_2$RuO$_4$. Our analysis seems quite general and also predicts
$d$-wave symmetry for electron- and hole-doped cuprates\cite{manske}.
To decide whether $p$- or $f$-wave symmetry pairing is present 
one needs to perform 
more complete calculations including coupling between RuO$_2$ planes, 
for example. If the interplane coupling involves also nesting, then 
corresponding nodes are expected.

We are thankful for stimulating discussions with P. Thalmeier, M. Sigrist,
R. Klemm, M. Kagan, and D. Agterberg. I.E. is 
supported by Alexander von Humboldt Foundation.


\begin{references}
\bibitem{matrix} We also include 
the matrix element for the spin susceptibility $\chi_0^{ij}$ 
taken to be 1 between two states which have the same 
maximal character and 0 otherwise in accordance with 
Ref. \protect\cite{mazin}.
\bibitem{remark} In the original paper\protect\cite{imai} the effect of
cross-susceptibilities was not considered. Therefore, we compare our results
to the total susceptibility $\chi_{xy}+\chi_{yz}+\chi_{xz}$.
\bibitem{remeuro} Since $\gamma$-band contains the information about the 
nesting properties of the $\alpha$- and $\beta$-bands and  due to the
neglection of the interband Coulomb interaction, we solve the gap equation 
only for the leading $\gamma$-band and taking into account 
the full pairing potential obtained 
from the dioganalizing of the [$\chi^{ij}$] matrix.
\bibitem{footbe} If the energy difference between $p$- and $f$-wave 
symmetry in the RuO$_2$ planes is larger than the energy gain for 
superconductivity resulting from interplane coupling, one may determine 
the pairing symmetry from the in-plane electronic structure.
\bibitem{remark2} Note, the second term in V$_0$ comes from ladder diagrams
which do not contribute to the triplet pairing due to the Pauli principle.
\end{references}
\end{document}